\documentclass[fleqn,usenatbib,useAMS]{mnras}

\usepackage{graphicx}	% Including figure files
\usepackage{amsmath}	% Advanced maths commands
\usepackage{amssymb}	% Extra maths symbols
\usepackage{multicol}        % Multi-column entries in tables
\usepackage{bm}		% Bold maths symbols, including upright Greek
\usepackage{pdflscape}	% Landscape pages
\usepackage{xtab}

\usepackage[T1]{fontenc}
\usepackage{ae,aecompl}

\usepackage{newtxtext,newtxmath}
\title[NGC 1052 halo mass]{Dark matter and no dark matter: On the halo mass of NGC 1052}
\author[D. A. Forbes et al.]{Duncan A. Forbes$^{1}$\thanks{Contact e-mail: \href{mailto:dforbes@swin.edu.au} dforbes@swin.edu.au}
Adebusola Alabi$^{2}$, Jean P. Brodie$^{2}$, Aaron J.  Romanowsky$^{2,3}$ 
\\
$^{1}$Centre for Astrophysics \& Supercomputing, Swinburne
University, Hawthorn VIC 3122, Australia\\
$^{2}$University of California Observatories, 1156 High Street, Santa Cruz, CA 95064, USA\\
$^{3}$Department of Physics and Astronomy, San Jos\'e State University, San Jos\'e, CA 95192, USA}

\date{Last updated 2015 May 22; in original form 2013 September 5}

\pubyear{2019}

\begin{document}
\label{firstpage}
\pagerange{\pageref{firstpage}--\pageref{lastpage}}
\maketitle

\begin{abstract}
The NGC 1052 group, and in particular the discovery of two ultra diffuse galaxies with very low internal velocity dispersions, has been the subject of much attention recently. Here we present radial velocities for a sample of 77 globular clusters associated with NGC 1052 obtained on the Keck telescope. Their mean velocity and velocity dispersion are consistent with that of the host galaxy. 
Using a simple tracer mass estimator, we infer the enclosed dynamical mass and dark matter fraction of NGC 1052. 
Extrapolating our measurements with an NFW mass profile we infer a total halo mass of 6.2 ($\pm$0.2) $\times$ 10$^{12}$ M$_{\odot}$. This mass is fully consistent with that expected from the stellar mass--halo mass relation, suggesting that NGC 1052 has a normal dark matter halo mass (i.e. it is not deficient in dark matter in contrast to 
two ultra diffuse galaxies in the group).
We present a phase space diagram showing the galaxies that lie within the projected virial radius (390 kpc) of NGC 1052. Finally, we briefly discuss the two dark matter deficient galaxies (NGC 1052--DF and DF4) and consider whether MOND can account for their low observed internal velocity dispersions. 

\end{abstract}


%%%%%%%%%%%%%%%%%%%%%%%%%%%%%%%%%%%%%%%%%%%%%%%%%%

%%%%%%%%%%%%%%%%% BODY OF PAPER %%%%%%%%%%%%%%%%%%

% The MNRAS class isn't designed to include a table of contents, but for this document one is useful.
% I therefore have to do some kludging to make it work without masses of blank space.
\begingroup
\let\clearpage\relax
%\tableofcontents
\endgroup
\newpage

\section{Introduction}

NGC 1052 is an elliptical galaxy with evidence of infalling HI gas and HI tidal tails (van Gorkom et al. 1986). The HI may have been acquired from the spiral galaxy NGC 1042 a few Gyr ago (van Gorkom et al.
1986). Dust lanes (Forbes et al. 1990) and misaligned star and gas kinematics (Davies \& Illingworth 1986) both indicate a recent past merger or interaction event. The current HI content, of a few 10$^8$ M$_{\odot}$ (van Gorkom et al. 1986), suggests that any contribution to the stellar mass from newly formed stars, or globular clusters (GCs), was fairly minor.
Indeed, the central galaxy is dominated by old ($\sim$ 13 Gyr) stars (Milone et al. 2007). It contains a low luminosity AGN with compact jets (Fern{\'a}ndez-Ontiveros et al. 2019). 
Forbes et al. (2001) found the GC system around NGC 1052 to be fairly  typical with a bimodal colour distribution and specific frequency S$_N$ = 3.2. The detected GCs have a range of luminosities, with the brightest being as luminous as $\omega$ Cen in the Milky Way. 
Spectra of 16 of the brighter GCs indicate typical old ages (Pierce et al. 2005). Recent deep imaging (Mueller et al. 2019) reveals stellar streams indicating an ongoing interaction between NGC 1052 and the S0 galaxy NGC 1047. 

NGC 1052, and its surrounding galaxies, have generated a great deal of interest lately with the claim of two relatively dark matter free galaxies, i.e. NGC 1052--DF2 and DF4 (van Dokkum et al. 2018, 2019). Some of the controversy has centred on whether DF2 and DF4 lie in the NGC 1052 group at $\sim$20 Mpc or much closer at $\sim$13 Mpc (Trujillo et al. 2019; Haghi et al. 2019); the closer distance would imply a more typical dark matter content. Also debated in the literature is whether DF2 is inconsistent with Modified Newtonian Dynamics (MOND) (van Dokkum et al. 2018) or consistent (Haghi et al. 2019). This debate centres around the extent of the external field effect (EFE) due to the mass of NGC 1052, the presence of which leads to a lower velocity dispersion prediction by MOND, in line with that observed for DF2 and DF4. 
The total mass of NGC 1052 is also key to understand whether tidal stripping can explain the lack of dark matter in these galaxies (Ogiya 2018; Nusser 2019).

%Thus it is of interest to measure the total mass of NGC 1052 and whether it is abnormal or not. 

A number of different methods can in principle be applied to estimate the dynamical mass of an individual galaxy (see review by Courteau et al. 2014). Here we focus on NGC 1052 and the kinematics of its GC system as tracers of the halo mass.
%which provides an estimate of the host galaxy mass from both the number of GCs and their motions. 
We wish to understand whether NGC 1052 has a typical dark matter halo or is deficient, as claimed for DF2 and DF4.

Assuming the standard Surface Brightness Fluctuation (SBF) calibration method is valid for NGC 1052, then its distance is relatively well-established, with four different SBF studies giving a range of 18.0 to 20.6 Mpc. Here we use a distance of 19.4 Mpc from the SBF study of Tonry et al. (2001). NGC 1052 is the brightest galaxy in a small group of galaxies known as LGG71 (Garcia 1993). 

In Table 1 we list some of the key properties of NGC 1052, e.g. it is an E4 elliptical with a predominately old stellar population, a modest central black hole
%a core-like surface brightness profile 
and a weak X-ray emitting diffuse halo. 
Its effective radius (R$_e$) has been measured as 21.9 arcsec in Forbes et al. (2017a) and 33.7 arcsec by Milone et al. (2007). For a distance of 19.4 Mpc, this translates to a range of 2.06 to 3.17 kpc. 

In this paper we present new radial velocities for a sample of GCs associated with NGC 1052. These GCs act as kinematic tracers of the dynamical mass out to the galactocentric radius of the outermost GC. We measure the dynamical mass (and dark matter fraction) and use it to infer the total halo mass. We compare this mass with that expected from the stellar mass-halo mass relation. 
Finally, we briefly discuss the cases of DF2 and DF4 in the context of MOND and whether they are bound to the NGC 1052 group.

\begin{table}
	\centering
	\caption{NGC 1052 Properties.}
	\label{tab:example_table}
	\begin{tabular}{lcrr} % four columns, alignment for each
		\hline
		Property & Value & Unit & Ref.\\
		\hline
		Type & E4 & & NED \\
		Velocity & 1510$\pm$6 & km/s & D05\\
		$\sigma$ & 176$\pm$26 & km/s & D05\\
		Distance & 19.4$\pm$2.6 & Mpc & T01\\
        Age & $\sim$13 & Gyr & M07\\
        R$_e$ & 2.1--3.2 & kpc & F17a, M07\\
        log M$_{\ast}$ & 11.02 & M$_{\odot}$ & F17a\\
        log M$_{BH}$ & 7.97 & M$_{\odot}$ & B09\\
%        SB profile & core & & K19\\
        log L$_X$ & 39.64$\pm$0.03 & erg/s & K19\\
		\hline
	\end{tabular}
\end{table}

\section{Observations and Data Reduction}

Spectroscopic observations of NGC 1052 GC candidates (Forbes et al. 2001) were obtained using the
DEIMOS spectrograph (Faber et al. 2003) on the Keck II 
telescope between 
between 2016 November and 2017 January.
The observations and data reduction follow that of the SLUGGS survey (Brodie et al. 2014; Forbes et al. 2017b), although NGC 1052 is not part of that survey. 
Briefly, the DEIMOS instrument is used in multi-slit mode with each
slit mask covering an area of $\sim$ 16 $\times$ 5 arcmin$^2$. 
We used the 1200 lines per mm setting, the OG550 filter, 
slit width of 1 arcsec and a central wavelength of
7800\AA ~(which includes the Calcium triplet features).
This setup gives a spectral resolution of $\sim$
1.5\AA ~(FWHM). 
Four slit-masks were obtained 
under seeing conditions of 0.6 to 1.3 arcsec. For observing details see 
Bellstedt et al. (2018) who analysed the slits near the galaxy centre  containing galaxy starlight. Here we use the slits dedicated to GC candidates of NGC 1052. 

The data have been reduced using the standard methods of the SLUGGS survey, e.g. Pota et al. (2013). This reduction process effectively removes the sky and galaxy background light from that of the GC. The radial velocity of each GC spectrum is determined using a set of a dozen stellar template stars observed with the same grating and central wavelength. Tests of repeatability (i.e. from observing the same mask
on different nights) from the SLUGGS survey 
indicates a systematic rms velocity resolution of
$\pm$10--15 km/s (Pota et al. 2013). 

\section{Results}

\subsection{Radial Velocities}

From the four DEIMOS masks we derive radial velocities from the Calcium Triplet lines for 77 GCs associated with NGC 1052 using the same method as described in Forbes et al. (2017b). The coordinates and radial velocity of each GC are given in Appendix A.

%{\bf Has substructure test been carried out?}

\begin{figure}
	\centering
	\includegraphics[angle=-90,width=1\linewidth]{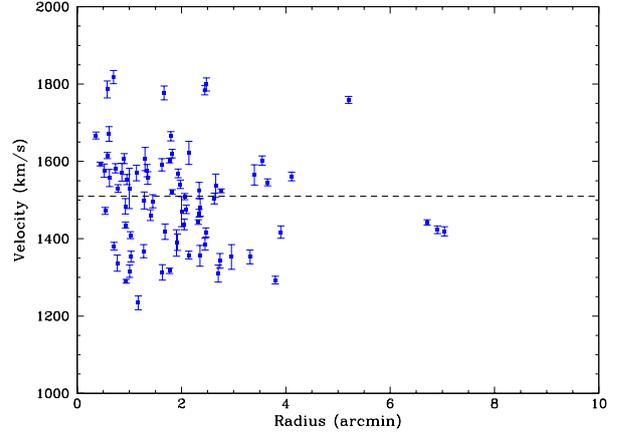}
	\caption{Phase space diagram for 77 GCs associated with NGC 1052 showing radial velocity vs projected galactocentric radius in arcsec. The horizontal line shows the systemic velocity of NGC 1052.}
	\label{fig:gcrv}
\end{figure}

In Fig. \ref{fig:gcrv} we show the phase space diagram for GCs associated with NGC 1052, i.e. their radial velocities as a function of projected galactocentric radius.  The GC system of NGC 1052 has a mean velocity of 1503 $\pm$ 15 km/s and a velocity dispersion of 132 $\pm$ 15 km/s. This is comparable to NGC 1052 itself with a systemic velocity of 1510 $\pm$ 6 km/s and a central velocity dispersion of 176 $\pm$ 26 km/s. 

\subsection{Dynamical Mass and Dark Matter Fraction}

Following the method outlined in Alabi et al. (2017) we apply the Tracer Mass Estimator (TME) of Watkins et al. (2010) to the 77 GCs associated with NGC 1052 (we do not use the 16 GCs of Pierce et al. 2005 as they have much higher velocity errors and are all located within the central regions of NGC 1052 where we have excellent coverage). 
The TME assumes a pressure-supported system in dynamical equilibrium (we find no evidence for substructure or strong rotation in the GC system).
We use a gravitational potential with slope $\alpha$ = 0.22 and a de-projected GC density slope of $\gamma$ = 2.9, i.e. typical values used in  equations 2 and 3 of Alabi et al. 
%We find little evidence of flattening in the GC distribution. 
Under the assumption of isotropic orbits the total mass within 5R$_e$ for the smaller R$_e$ value (i.e. 5 $\times$ 2.06 kpc) is 1.17 ($\pm$ 0.26) $\times$ 10$^{11}$ M$_{\odot}$. 
%Removing the rotational contribution to this mass would only decrease it by $\le$10\%. 
Relaxing the isotropic assumption, to include mild radial or tangential orbits (as per Alabi et al. 2017), results in masses that are within $\pm$4\% of the mass for the isotropic case. 
%The main source of uncertainty is the effective radius. 
If we adopt the larger R$_e$ value of 3.2 kpc, then the dynamical mass within 5R$_e$ becomes 
1.71 ($\pm$ 0.32) $\times$ 10$^{11}$ M$_{\odot}$. 

%{\bf drop this para?}\\
%The dark matter fraction within 5R$_e$ is also subject to large change if one uses R$_e$ = 2.06 kpc or 3.2 kpc, i.e. f$_{DM}$ = 0.18 $\pm$ 0.26 or 0.44 $\pm$ 0.14. Alabi et al. (2017) derived f$_{DM}$ within 5R$_e$ using GCs for a sample of early-type galaxies as part of the SLUGGS survey. They found a very wide range in dark matter fractions from 0.1 to 0.9 for galaxies with a stellar mass similar to NGC 1052. Some of this variation may be due to uncertainties in the measurement of the host galaxy effective radii. 

van Gorkom et al. (1986) noted that the HI gas in NGC 1052 appeared to be rotating with disk-like kinematics but also showed some evidence of warping. Under the assumption of ordered motions, and 
scaling to our distance, they derived a dynamical mass from the rotation velocity of the HI gas within a radius of 23 kpc of 3.1 $\times$ 10$^{11}$ M$_{\odot}$. 
%orginal was 1.5x10(11) within 16 kpc for D=13.4Mpc.
We have also calculated our mass within a physical radius of 23 kpc and find 
2.44 ($\pm$ 0.44) $\times$ 10$^{11}$ M$_{\odot}$ for an isotropic orbit, i.e. within 2$\sigma$ of the HI derived mass. This difference in mass would be reduced if the GCs at large radii are on more radial orbits (which we can not constrain with our data).

The maximum radius reached by our GCs is 40 kpc (or $\sim$13--20 galaxy effective radii). 
Within this radius we calculate a total mass of 
4.14 ($\pm$ 0.71) $\times$ 10$^{11}$ M$_{\odot}$ and a dark matter fraction of f$_{DM}$ = 0.75 $\pm$ 0.06. The uncertainty on the dark matter fraction is much reduced at this radius compared to that at 5R$_e$ and it is insensitive to the galaxy's actual effective radius. We find the galaxy to be strongly dark matter dominated at this radius, similar to the early-type galaxies of the SLUGGS survey (Alabi et al. 2017). 
%In the next subsection we use this outer radius enclosed mass to estimate the total halo mass of the galaxy. 

\subsection{Halo Mass}

Under the assumption that the mass in the halo follows a standard NFW profile (Navarro et al. 1997) under a $\Lambda$CDM cosmology, we extrapolate the enclosed mass within 5R$_e$ to the virial radius to estimate a total halo mass of 6.2 ($\pm$0.2) $\times$ 10$^{12}$ M$_{\odot}$ or log M$_h$ = 12.79 $\pm$ 0.02 for a concentration parameter c = 7.0.
This halo mass  and its quoted uncertainty is the average of using the two different effective radii (see Table 1). Thus the derived halo mass is relatively insensitive to the adopted effective radius of the galaxy (but will vary for different adopted mass profiles or halo 
concentration parameters etc). 
The virial radius corresponding to this halo mass is 390 kpc (assuming cosmological parameters from  Planck collaboration et al. 2018).

%12.77 for Re=2.06, 12.81 for Re=3.17

%An alternative method for estimating the halo mass comes from counts of the total number of GCs (Burkert \& Forbes 2019) and the total mass of a GC system (Forbes et al. 2018). Both have a strong near-linear relation with the host galaxy halo mass. Here we use the formula given in Burkert \& Forbes (2019) and the number of GCs associated with NGC 1052 from Forbes et al. (2001), i.e. 400, to estimate a halo mass of log M$_h$ = 12.31 
%$\pm$ 0.05.

%Kim et al. (2019) give the relation between L$_X$ luminosity of the halo and the dynamical mass within 5R$_e$. Here we taken the relation for core galaxies and the X-ray luminosity for NGC 1052 (see Table 1) to estimate log M$_{5Re}$ = 11.41
%LX  = 2.36 M5Re + 12.71
%Assuming an NFW mass profile this gives a halo mass of ***

In order to determine whether this halo mass is as expected for such a galaxy, we compare it with the expectation from the  stellar mass--halo mass relation (SMHR). Although the SMHR is not well determined in the dwarf galaxy (and perhaps the brightest cluster galaxy) regime it is quite well determined, with good agreement between different authors, for galaxies around L$^{\ast}$ such as NGC 1052. The stellar mass for NGC 1052 is taken from Forbes et al. (2017a) which is based on the 
3.6$\mu$ luminosity of NGC 1052 and an appropriate mass-to-light ratio for an old stellar population.
%In Fig. \ref{fig:smhr} we show the inverted halo mass--stellar mass relation from the work of

Using the inverted SMHR relation of Rodriguez-Puebla et al. (2017) and a log stellar mass of 11.02, a log halo mass of 12.76 is predicted. We note that the last term in their eq. 66 has a sign error and should be -1/2 (Rodriguez-Puebla, priv. comm.).
The typical scatter about the relation in this mass regime is $\sim$0.2 dex. This is fully consistent 
with our estimate of the halo mass (i.e. 12.79 $\pm$ 0.02), scaled from our 5R$_e$ dynamical mass. Given the assumptions above, we conclude that NGC 1052 has a typical dark matter halo.

%\begin{figure}
%	\centering
%	\includegraphics[angle=-90,w%idth=1\linewidth]{smhr.pdf}
%	\caption{Halo mass--stellar %mass relation. The solid line %represents the halo %mass-stellar mass relation of %Rodriguez-Puebla et al. (2017), %which has an intrinsic scatter %of around $\pm$0.2 dex. Our %estimated halo mass for NGC %1052 is perfectly consistent %with its measured stellar mass. %}
%	\label{fig:smhr}
%\end{figure}

%The stellar mass--halo mass relation can also be used to estimate the halo mass of NGC 1052. Although the relations differ in the low mass regime, they are good agreement at higher masses of massive elliptical galaxies like NGC 1052. Here we use the relation of Hudson et al. (2015) 
%logM200 = 0.57 log M* + 5.97 from email with Mike
% and the stellar mass of NGC 1052 (see Table 1) to estimate a halo mass of log M$_h$ = 12.25. 

\section{Discussion}

Based on our dynamical measurements, and an assumption of a  $\Lambda$CDM cosmology, we have estimated a total halo mass for NGC 1052. We find that it is dark matter dominated and its halo mass is completely consistent with that expected based on its measured stellar mass. 
%Based on its halo mass, the inferred virial radius for the dark matter halo is 390 kpc (or 70 arcmin). 
We note that in a MONDian universe, the dark matter mass is a phantom mass (Wu \& Kroupa 2015) and the total mass of the galaxy is simply the baryonic mass (i.e. stellar plus gas mass).

\begin{figure}
	\centering
	\includegraphics[angle=-90,width=1\linewidth]{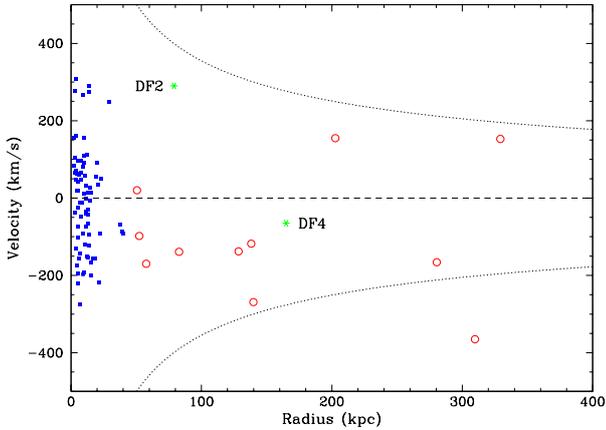}
	\caption{Phase space diagram of objects around NGC 1052
	showing radial velocity corrected for the systemic velocity of NGC 1052 vs projected galactocentric radius in kpc assuming a distance of 19.4 Mpc. Filled blue circles are GCs, open red circles are galaxies and green asterisks are UDGs DF2 and DF4. The horizontal dashed line shows the systemic velocity of NGC 1052, and the dotted lines show the velocity caustics corresponding to the dark matter halo of NGC 1052 in a logarithmic potential.}
	%for a 6.2 $\times$ 10$^{12}$ M$_{\odot}$ point mass.}
	\label{fig:grouprv}
\end{figure}

In Fig.~\ref{fig:grouprv} we extend our phase space diagram out to the projected virial radius of 390 kpc and subtract the systemic velocity of NGC 1052 (1510 km/s) with the observed line-of-sight radial velocities. As well as showing the GCs, we include galaxies listed by NED (e.g. NGC 1047 r=57 kpc, V=1340 km/s; NGC 1042 r=83 kpc, V=1371 km/s; NGC 1035 r=140 kpc, V = 1241 km/s) 
and the Ultra Diffuse Galaxies (UDGs) DF2 and DF4. We also show velocity caustics corresponding to the halo mass of NGC 1052 
%a log halo mass = 12.79 M$_{\odot}$ 
in a logarithmic potential. Here the circular velocity of the halo is 259 km/s, the halo scale radius is 41.5 kpc and concentration parameter c = 7). 
According to van Dokkum et al. (2019) the NGC 1052 group has a mean velocity of V = 1438 $\pm$ 25 km/s and a velocity dispersion of 128 $\pm$ 19 km/s.
The latter is quite consistent with our measurement of the GC system, i.e. 132 km/s.

From Fig.~\ref{fig:grouprv} 
it is likely that DF4 is probably bound to the NGC 1052 group. This is less clear for DF2 (also known as KKS[2000]04) which has a much higher recession velocity (although it is similar to the highest velocity GCs we observe) and may have a true 3D separation closer to $\sqrt{3/2}$ times greater than its projected radius. Its position in phase space is also suggestive of being on an initial infall into the group (Rhee et al. 2017).

DF2 has been observed to have a very low velocity dispersion (e.g. $\sim$10 km/s depending on how errors and small sample size are handled) for both its stars and its GC system (van Dokkum et al. 2018). Under the assumption that it lies at the distance of NGC 1052 (i.e. ~$\sim$20 Mpc) this implies a total mass, within the observed radius,  with little or no dark matter. 
%In MONDian theory, dark matter is not physical but is rather a psuedo mass inferred by applying Newtonian gravity. 
It was argued by van Dokkum et al. (2018) that the very low velocity dispersion measured in DF2 could not be explained by MOND.
This would be true if DF2 was isolated. 
However, the velocity dispersion measured for DF2 could be influenced by the External Field Effect (EFE) if DF2 resides in a large potential, i.e. it will be lower than that for an isolated system (e.g. McGaugh \& Milgrom 2013). 
%For example, MOND has had some success in reproducing the low velocity dispersions for satellites embedded in the Milky Way's halo (McGaugh \& Milgrom 2013). 

Recently, several authors have predicted the velocity dispersion for DF2 (and other UDGs in the NGC 1052 group) based on MOND and the EFE influence of NGC 1052 (Famaey et al. 2018; Kroupa et al. 2019; Haghi et al. 2019; Mueller et al. 2019). These works show that the observed internal velocity dispersion of DF2 does not rule out MOND. 
Haghi et al. (2019) also explored the EFE if both DF2 and NGC 1052 are at distance of $\sim$10 Mpc, and concluded that this would make the MOND-predicted velocity dispersion fully consistent with the observations. 
However, unless one invokes an abnormal stellar population, the distance of NGC 1052 is quite secure with multiple SBF measurements all giving a distance closer to 20 Mpc. 
Kroupa et al. (2019) point out that MOND can be falsified if DF2 is at a distance of greater than 18 Mpc {\it and} more than 300 kpc from the nearest large galaxy. 
%If DF2 were at a distance of 10 Mpc {\it and} was isolated from NGC 1052 (and other large galaxies) then the observed velocity dispersion would not be consistent with MOND (Haghi et al. 2019). 

DF4 has also been found to have a very low velocity dispersion based on its GC system (van Dokkum et al. 2019). 
Although located at a projected radius of 165 kpc from NGC 1052, Fig. \ref{fig:grouprv} suggests it is bound to the NGC 1052 group and hence at a distance of 20 Mpc. Haghi et al. (2019) showed that the EFE from NGC 1052 is able to reproduce the  observed velocity dispersion but only within the 3$\sigma$ level.  However, the edge-on spiral galaxy NGC 1035 lies only 20 kpc away in projection. In this case, the EFE of NGC 1035 is much closer to reproducing the observed velocity dispersion.
%if the projected separation is close to the 3D one.
%if the true separation is $\le$10 kpc. 

Although the distance to NGC 1052 (and its associated GCs and  group galaxies) is secure at $\sim$20 Mpc we
acknowledge the considerable debate in the literature on the true distances to DF2 and DF4. For example,  Cohen et al. (2018) argued for a distance of $\sim$20 Mpc, while 
Monelli 
\& Trujillo (2019) argued that DF2 and  DF4 (along with NGC 1035 and NGC 1042)  lie at $\sim$13 Mpc. More work is needed to understand and verify these different distance determinations.
%{\bf Recently, Haslbauer et al. %(2019) have shown that DF2 and %DF4-like 
%analogues, with little dark matter %(corresponding to a distance of %$\sim$20 Mpc), are extremely rare %in $\Lambda$CDM cosmological %simulations.} 
Another uncertainty, which will remain, is the true 3D separation between DF2 and DF4 and the larger galaxies. This determines the relative isolation of each galaxy and therefore  the strength of the EFE  in MOND, and as well as the strength of any tidal interactions. 
%while NGC 1052 and NGC 1047 is at $\sim$20 Mpc. 

\section{Conclusions}

Here we present radial velocities for 77 globular clusters associated with the brightest group elliptical NGC 1052. Data were obtained using the DEIMOS spectrograph on the Keck II telescope. We measure a GC system mean velocity of 1503 km/s and a velocity dispersion of 132 km/s. 
These values are consistent with that of NGC 1052 itself. Using the GCs as tracer particles we estimate a total dynamical mass within 40 kpc of 4.14 $\times$ 10$^{11}$ M$_{\odot}$ 
(compared to a stellar mass of $\sim$10$^{11}$ M$_{\odot}$) and hence a high dark matter fraction of 75\%. Extrapolating our measurements to the virial radius, assuming an NFW profile, we estimate a halo mass of 6.2 ($\pm$0.2) $\times$ 10$^{12}$ M$_{\odot}$. This mass is fully consistent with that expected from the stellar mass--halo mass relation, suggesting that NGC 1052 hosts a typical, large dark matter halo.  
Based on a phase space diagram out to the projected virial radius of NGC 1052, we suggest that the ultra diffuse galaxy DF4 is most likely bound to the group and hence lies at $\sim$20 Mpc, whereas this is less clear for DF2. The proximity of DF2 and DF4 to a larger host galaxy like NGC 1052 has implications for whether MONDian gravity can explain the low observed velocity dispersions in these two galaxies. Their true distances, and hence their membership of the NGC 1052 group, is still debated.

\section{Acknowledgements}

We thank the referee for their comments. 
JPB and AJR acknowledge NSF grant AST-1616598. 
AJR was supported by National Science Foundation grant AST-1616710 and as a Research Corporation for Science Advancement Cottrell Scholar. We thank A. Ferre-Mateu, A. Wasserman, V. Pandya, L. Chevalier and S. Bellstedt for their help with the mask  design and/or observations. We thank P. van Dokkum, P. Kroupa and H. Haghi for their comments on an early version of the paper. 
The data presented herein were obtained at the W. M. Keck Observatory, which is operated as a scientific partnership among the California Institute of Technology, the University of California and the National Aeronautics and Space Administration. The Observatory was made possible by the generous financial support of the W. M. Keck Foundation. The authors wish to recognize and acknowledge the very significant cultural role and reverence that the summit of Maunakea has always had within the indigenous Hawaiian community.  We are most fortunate to have the opportunity to conduct observations from this mountain.

\section{References}

Alabi A.~B., et al., 2017, MNRAS, 468, 3949 \\
Beifiori A., Sarzi M., Corsini E.~M., Dalla Bont{\`a} E., Pizzella A., Coccato L., Bertola F., 2009, ApJ, 692, 856 (B09)\\
Bellstedt S., et al., 2018, MNRAS, 476, 4543 \\
%Bird S., Harris W.~E., Blakeslee J.~P., Flynn C., 2010, A\&A, 524, A71 \\
Brodie J.~P., et al., 2014, ApJ, 796, 52 \\
Cohen Y., et al., 2018, ApJ, 868, 96 \\
Courteau S., et al., 2014, RvMP, 86, 47 \\
Davies R.~L., Illingworth G.~D., 1986, ApJ, 302, 234 \\
Denicol{\'o} G., Terlevich R., Terlevich E., Forbes D.~A., Terlevich A., 2005, MNRAS, 358, 813 (D05)\\
Faber S.~M., et al., 2003, SPIE, 4841, 1657 \\
Famaey B., McGaugh S., Milgrom M., 2018, MNRAS, 480, 473 \\
Fern{\'a}ndez-Ontiveros J.~A., L{\'o}pez-Gonzaga N., Prieto M.~A., Acosta-Pulido J.~A., Lopez-Rodriguez E., Asmus D., Tristram K.~R.~W., 2019, MNRAS, 485, 5377 \\
Garcia A.~M., 1993, A\&AS, 100, 47 \\
%Haslbauer, M., et al. 2019, MNRAS, accepted\\
Forbes D.~A., Sparks W.~B., Macchetto F.~D., 1990, NASCP, 3098\\ 
Forbes D.~A., Georgakakis A.~E., Brodie J.~P., 2001, MNRAS, 325, 1431 \\
Forbes D.~A., Sinpetru L., Savorgnan G., Romanowsky A.~J., Usher C., Brodie J., 2017a, MNRAS, 464, 4611 (F17a)\\
Forbes D.~A., et al., 2017b, AJ, 153, 114 \\
%Haghi H., Bazkiaei A.~E., Zonoozi A.~H., Kroupa P., 2016, MNRAS, 458, 4172 \\
Haghi H., et al., 2019, MNRAS, 487, 2441 \\
Kim, D-W., et al. 2019, MNRAS, in press (K19)\\
Kroupa P., et al., 2019, arXiv, arXiv:1903.11612 \\
Milone A.~D.~C., Rickes M.~G., Pastoriza M.~G., 2007, A\&A, 469, 89 (M07)\\
Monelli M., Trujillo I., 2019, arXiv, arXiv:1907.03761 \\
McGaugh S., Milgrom M., 2013, ApJ, 775, 139 \\
M{\"u}ller O., et al., 2019, A\&A, 624, L6 \\
Navarro J.~F., Frenk C.~S., White S.~D.~M., 1997, ApJ, 490, 493 \\
Nusser A., 2019, arXiv, arXiv:1907.08035 \\
Ogiya G., 2018, MNRAS, 480, L106 \\
Pierce M., Brodie J.~P., Forbes D.~A., Beasley M.~A., Proctor R., Strader J., 2005, MNRAS, 358, 419 \\
Pota V., et al., 2013, MNRAS, 428, 389 \\
Planck Collaboration, et al., 2018, arXiv, arXiv:1807.06209 \\
Rhee J., Smith R., Choi H., Yi S.~K., Jaff{\'e} Y., Candlish G., S{\'a}nchez-J{\'a}nssen R., 2017, ApJ, 843, 128 \\
Rodr{\'{\i}}guez-Puebla A., Primack J.~R., Avila-Reese V., Faber S.~M., 2017, MNRAS, 470, 651 \\
Tonry J.~L., Dressler A., Blakeslee J.~P., Ajhar E.~A., Fletcher A.~B., Luppino G.~A., Metzger M.~R., Moore C.~B., 2001, ApJ, 546, 681 (T01)\\
Trujillo I., et al., 2019, MNRAS, 486, 1192 \\
%van Dokkum P., et al., 2018, Natur, 555, 629 \\
van Dokkum P., Danieli S., Cohen Y., Romanowsky A.~J., Conroy C., 2018, ApJ, 864, L18 \\
van Dokkum P., Danieli S., Abraham R., Conroy C., Romanowsky A.~J., 2019, ApJ, 874, L5 \\
van Gorkom J.~H., Knapp G.~R., Raimond E., Faber S.~M., Gallagher J.~S., 1986, AJ, 91, 791 \\
Watkins L.~L., Evans N.~W., An J.~H., 2010, MNRAS, 406, 264 \\
Wu X., Kroupa P., 2015, MNRAS, 446, 330 \\

%%%%%%%%%%%%%%%%%%%%%%%%%%%%%%%%%%%%%%%%%%%%%%%%%%

\newpage

\section{Appendix}

In this Appendix we list the measured radial velocities from Keck/DEIMOS spectra for 77 globular clusters associated with NGC 1052. \\

\bigskip

%\tablefirsthead{}
%\bigskip

%\topcaption{.~Globular Cluster Radial Velocities.}

%First&\multicolumn{1}{c}{Name} \\ \midrule}
%

%\tablehead{%
%\multicolumn{5}{c}%
%{{\bfseries  }} \\
%\toprule Continued...\\}

%&\multicolumn{1}{c}{bbb}\\ \midrule}
%
%\tabletail{%
%\midrule \multicolumn{5}{c}{} \\}
%\tablelasttail{%
%\\\midrule
%\multicolumn{2}{r}{} \\}

\begin{tabular}{lllll}

%	\caption{Globular Cluster Radial Velocities.}
%	\label{tab:example_table}
%	\begin{tabular}{lcccc} 

\hline
		ID & RA & Dec. & Velocity & Error\\
		   & (J2000) & (J2000) & (km/s) & (km/s)\\
\hline
  NGC1052$\_$GC1  &40.249925 &-8.245486 &1557 &16    \\
  NGC1052$\_$GC3  &40.255600   &-8.251517 &1607 &13    \\
  NGC1052$\_$GC4  &40.294421 &-8.288517 &1386 &15    \\
  NGC1052$\_$GC5  &40.232761 &-8.230366 &1310 &22    \\
  ... & ... & ... & ... & ...\\
\end{tabular}

\bigskip

% Don't change these lines
\bsp	% typesetting comment
\label{lastpage}
\end{document}